# Architecture Information Communication in Two OSS Projects: the Why, Who, When, and What

**Tingting Bi** [a,b], **Wei Ding** [e,a], **Peng Liang** [a,*], **Antony Tang** [c,d]

[a] School of Computer Science, Wuhan University, 430072 Wuhan, China
[b] Faculty of Information and Technology, Monash University, VIC 3166, Melbourne, Australia
[c] Faculty of Science, Engineering and Technology, Swinburne University of Technology, VIC 3122 Melbourne, Australia
[d] Software and Services Research Group, Vrije Universiteit Amsterdam, 1101, Amsterdam, The Netherlands
[e] Key Laboratory of Earthquake Geodesy, Institute of Seismology, China Earthquake Administration, China
bi_tingting@whu.edu.cn, tingwhere@whu.edu.cn, liangp@whu.edu.cn, atang@swin.edu.au

## Abstract

Architecture information is vital for Open Source Software (OSS) development, and mailing list is one of the widely used channels for developers to share and communicate architecture information. This work investigates the nature of architecture information communication (i.e., why, who, when, and what) by OSS developers via developer mailing lists. We employed a multiple case study approach to extract and analyze the architecture information communication from the developer mailing lists of two OSS projects, ArgoUML and Hibernate, during their development life-cycle of over 18 years. Our main findings are: (a) architecture negotiation and interpretation are the two main reasons (i.e., why) of architecture communication; (b) the amount of architecture information communicated in developer mailing lists decreases after the first stable release (i.e., when); (c) architecture communications centered around a few core developers (i.e., who); (d) and the most frequently communicated architecture elements (i.e., what) are Architecture Rationale and Architecture Model. There are a few similarities of architecture communication between the two OSS projects. Such similarities point to how OSS developers naturally gravitate towards the four aspects of architecture communication in OSS development.

[*] Corresponding author at: School of Computer Science, Wuhan University, China.
   Tel.: +86 27 68776137; fax: +86 27 68776027. E-mail address: liangp@whu.edu.cn (P. Liang).

# 1. Introduction

Software Architecture (SA) of a system is "*the set of structures needed to reason about the system, which comprise software elements, relations among them, and properties of both*" [7]. In traditional software development, SA documentation is an important source used for communicating and sharing architectural information [11]. But our recent survey shows that SA documentation has not been widely practiced in Open Source Software (OSS) development (only 5.4% of 2000 popular OSS projects have SA documentation) [16]. Yet, software architectural information is relevant for OSS as well; in OSS, SA is used to promote anarchic collaboration yet preserve centralized control [18]. How then do developers in OSS development communicate and share architectural information? Participatory culture is prevalent in software development to share ideas and knowledge about requirements, architecture, code, etc., especially in the OSS community where developers intensively communicate in a distributed environment, and social media plays an essential role in this culture [42].

Communication and documentation are vital in software development, and various channels are used for this purpose. Formal documentation such as architecture design specification captures design of projects. Informal documentation, such as issue tracker, Wiki, Slack, and email, captures design conversations and facilitates a distributed working environment, allowing problem discussions and task assignment to take place [59]. Research works show that mailing lists provide an important perspective to understanding communication between developers in OSS development, especially during software development life cycle [22][55][56][57][58][60].

Mailing lists have been used to record OSS project information, providing rich architecture and development information [22]. Mailing list is a communication tool used by developers since the beginning of the two OSS projects that we investigate, ArgoUML and Hibernate. The developer mailing list of ArgoUML has over 20 years of communication history from 2000 to 2020. Hibernate has over 18 years of communication history from 2002 to now. With a long history of developers' communication provided in these mailing lists, the goal of this study is to understand architecture information communication in OSS development.

Mailing lists in OSS development have been investigated recently for traceability between emails and source code [5], communication in development using mailing lists [22], and information seeking through mailing lists [39]. For example, a recent study on 37 Apache projects shows that 89.51% of all design discussions occur in project mailing lists [62]. However, to the best of our knowledge, there are no studies that investigate on the nature of architecture information communication using mailing lists. The **goal of this work** is to understand architecture information communication in OSS development using developer mailing lists: **Why** is architecture information communicated? **Who** communicates architecture information in OSS development? **When** do participants communicate architecture information? **What** architecture information is communicated during development? Answering these questions helps to understand the characteristics of such communication, and provides a basis for evaluating best practices of OSS architecture information communication and development.

We investigate architecture information communication in the developer mailing lists of two OSS projects:

ArgoUML and Hibernate. We also checked other channels that developers used to communicate development knowledge in these two OSS projects, and found that developer mailing lists are mainly used by developers spanning the whole development life cycle (see Section 3.3). We manually checked all the posts of the developer mailing lists of these two projects, from the establishment of the mailing lists in Jan 2000 (ArgoUML) and Jan 2002 (Hibernate) to July 2020. We extracted in the mailing lists architectural posts (i.e., a single post that contains architectural information) and architectural threads (i.e., a sequence of posts communicated on the same conversation topic that contain architectural information). One thread may contain only one post if the topic introduced does not result in any replies. We identified 316 architectural posts and 200 architectural threads from 26,647 posts in the developer mailing list of ArgoUML, and 257 architectural posts and 159 architectural threads from 22,888 posts in the developer mailing list of Hibernate.

Analyzing these architectural posts and threads, we have the following findings: (1) the main purpose of architecture information communication in OSS developer mailing lists is to discuss architecture issues and questions; (2) a few core developers communicated architecture information intensively, and played the role of architect in OSS development; (3) architecture information communicated in OSS mailing lists was intense before the first stable releases and decreased afterwards; (4) amongst the architecture elements in the ISO 42010:2011 standard [25], the most frequently communicated architecture elements in OSS developer mailing lists are Architecture Rationale, Architecture Model, and Concern. Four topics of architecture information communication content have been found: architecture topic, abstraction level, internal quality requirement, and external quality requirement. These findings have been analyzed and explained with their implications in OSS development in this paper.

This paper is organized as follows: Section 2 introduces related work on the use of mailing lists and architecture information communication in OSS projects. Section 3 describes the case study design. The results of the case study are presented in Section 4, and the implications of the results are discussed in Section 5. Section 6 discusses the threats to validity, and Section 7 concludes this work with future directions.

## 2. Related work

Mailing lists have been intensively investigated in OSS development. Sowe *et al.* analyzed the mailing lists of the Debian project, and intended to understand the knowledge sharing process in OSS projects between knowledge providers and seekers [41]. Jensen *et al.* investigated how the first posts of newcomers were reacted in the mailing lists of four OSS projects: MediaWiki, GIMP, PostgreSQL, and Subversion. The results show that 80% of these posts got prompt feedback within two days, and this timely response is positively correlated with the future participation of newcomers in the OSS community [27]. Mockus *et al.* analyzed code change history with issue reports archived in mailing lists to present quantitatively various aspects of the development, e.g., developer participation and issue resolution intervals, for two successful OSS projects Apache and Mozilla . These works used posts in the mailing lists of an OSS project without specifically extracting and analyzing architectural information in mailing lists.

Information available in OSS communities is used to investigate the communication methods in OSS development. For example, Avritzer *et al.* used design structure matrix to represent an architectural structure as design decision and to assess the communication structure of developers in OSS projects GSP 3.0 [4]. They used a questionnaire to collect social interactions between the development teams. Crowston *et al.* examined the bug tracking systems of various OSS projects and revealed that the communication centralizations of various OSS project teams are different [13]. Guzzi *et al.* used an OSS project Lucene and sampled 506 threads from the developer mailing list [22]. Their work provides a relative frequency of topics communicated and participation of developers in the mailing list. They paid attention to the communication of all posts and threads in mailing lists, which is not restricted to architectural posts and threads in mailing lists. Bass *et al*. investigated whether a formal architecture document can help to communicate architecture knowledge better and influence OSS development. They found that it helps inexperienced developers to progress faster [8]. Brunet *et al.* developed a classifier to automatically label 102,122 discussions (i.e., a set of comments on pull requests, commits, and issues in GitHub) extracted from 77 OSS projects, and they observed that 25% of the discussions are about design [10], which is one of the major sources of architectural knowledge.

Ali *et al.* conducted a literature review on Architecture Knowledge Management (AKM) in Global Software Development (GSD) [2]. They defined a metamodel for the central concepts of AKM in GSD identified from the literature. The core concept of architecture information communication in this metamodel is *Coordination Strategies*, which is used to communicate *Design Decisions* to *Distributed Teams* in a distributed environment, and one practice of architecture knowledge *Coordination Strategies* in GSD is having a mailing list to quickly get information. The perceived usefulness of mailing lists for AKM in GSD steadily increases when more sites are involved in a software project [12]. According to our [16] survey on architecture documentation in OSS development, few (5.4%) OSS projects documented architecture knowledge, and in these OSS architecture documents, rationale (including design decisions) was rarely documented (18.5%). Investigating mailing lists, as a coordination strategy practice of architecture knowledge in distributed development, can help to understand the communication and sharing of architecture knowledge in OSS development.

In our previous work [48], we investigated the relationships between three specific design elements (i.e., architecture patterns, quality attributes, and design contexts) from Stack Overflow. Specifically, we analyzed how developers use architecture patterns with respect to quality attributes and design contexts, and uncovered relationships between these design elements. In another work [49], we extracted posts related to architecture smells from Stack Overflow, and we analyzed the extracted posts about developers' descriptions of architecture smells. We found that architecture smells are mainly caused by violating architecture patterns, design principles, or misusing architecture antipatterns. We further developed a semi-automatic approach to mine Architecture Tactic (AT) and Quality Attribute (QA) knowledge from Stack Overflow posts [63], and we then manually analyzed the mined posts for structuring the design relationships between ATs and QAs in practice. Such knowledge can help architects to better make design decisions on architecture tactics by considering quality attributes. These previous works also focus on architecture information extraction and analysis, while in this study, we aim at understanding architecture information communication through developer mailing lists.

## 3. Research design: multiple case study

To explore architecture information communication in OSS projects using mailing lists, we employ a multiple case study approach to analyze the developer mailing lists of two OSS projects: **ArgoUML** and **Hibernate** by following the guidelines in [36][43]. We limit the scope to developer mailing lists [22] without including e.g., user mailing lists, because developer mailing lists contain more development information (e.g., about architecture) than other types of mailing lists. A case study is suitable to provide an understanding of contemporary phenomena, such as the correlation between mailing lists and architecture information, which are hard to study in isolation. We employed a descriptive case study [36] in this work to portray a situation and phenomenon in architecture information communication using OSS developer mailing lists. The design of this case study is elaborated as follows: the objective and research questions are presented in Section 3.1, the definitions of case and units of analysis are described in Section 3.2, the criteria for case selection and the selected cases are presented in Section 3.3, and the processes of data collection and data analysis are detailed in Section 3.4 and Section 3.5 respectively. We selected posts by referencing the ISO 42010:2011 standard [25] to identify and collect architectural posts (described in Section 3.4). We employed certain techniques in Grounded Theory to extract and analyze the data items (D2 and D4) as detailed in Section 3.5.

## 3.1 Objective and research questions

The goal of this work is to understand the characteristics (i.e., the Why, Who, When, and What) of architecture information communication in OSS development life cycle, and we particularly targeted developer mailing lists as the communication channel. We chose to analyze textual information in developer mailing lists of two OSS projects (i.e., ArgoUML and Hibernate), and the context of the analysis is OSS development. We come up with four Research Questions (RQs) according to the goal. These RQs are shown in Table 1 together with their rationale. The answers to the four RQs can be directly mapped to the goal of this study: a practical understanding of the purposes of architecture information communication using mailing lists (RQ1), the developer social network on architecture information communication (RQ2), architecture information communication frequency in a time perspective (RQ3), and detailed content of architecture information communicated in mailing lists in a time perspective (RQ4).

**Table 1.** Research questions of this case study

| Research Question | Rationale |
|---|---|
| **RQ1**: What are the purposes (why) of communicating architecture information using mailing lists? | Developers communicate architecture information in mailing lists for different purposes, e.g., interpretation, negotiation, and rationale of architecture design, and different social media can be specifically used for different purposes of architecture information communication, for example, twitter is used to announce information, and blogs are used to explain the technical details of an architecture design. By investigating this RQ, we intend to understand the main purposes, or why, of communicating architecture information through mailing lists by OSS developers. This could help practitioners to better use mailing lists for specific purposes in architecture information communication. |
| **RQ2**: Who communicate architecture information in OSS projects using mailing lists from a social network perspective? | Mailing lists, as a type of social media, have been extensively employed to share knowledge between developers in OSS development [22]. The social network of participants on mailing lists has been used to study communication and coordination (C&C) activities in OSS development [9]. By answering this RQ, we aim to understand the communicating architecture and identify the form of C&C network of developers when they use mailing lists to communicate architecture information. For example, who is leading and contributing to the architecture conversations and how many developers are involved in the conversations. This information may help identify potential issues and improvements of architecture information communication in OSS development. |
| **RQ3**: When do OSS developers use mailing lists to communicate (e.g., *mention*, *ask*, and *reply*) architecture information? | The posts of mailing lists could be broadcasted and received by individuals who need the information (e.g., notification posts), or who possess the knowledge to answer the questions in the posts [21] (e.g., query posts). The architecture information discussed in mailing lists is scattered in threads of mailing lists and can be communicated at any time during development. By answering this RQ, we intend to identify the development activities when architecture information is communicated. This would allow us to investigate the relationship between frequency of architecture information communication and specific development activities, e.g., architecture change [31]. |
| **RQ4**: What architecture information is communicated in OSS development? | Architecture information is discussed in mailing lists according to different needs of participants. Developers may also communicate different information at different phases of OSS development, e.g., consider quality requirements in an initial stage of development. By answering this RQ, we intend to identify and classify the content of architecture information communication in mailing lists during OSS development, which is meaningful for researchers and practitioners to understand and improve the way architecture information is discussed and constructed in OSS development. The conceptual model for architectural description in the ISO 42010:2011 standard [25] is used to structure the communicated architectural information in mailing lists. |

## 3.2 Case and units of analysis

The cases used in this case study are the two OSS projects, and the units of analysis are architectural posts and architectural threads of the OSS projects. They are the basic elements in this multiple case study [36].

## 3.3 Case selection

To ensure that the cases (OSS projects) investigated in this study are non-trivial and representative, we define three criteria for case selection: (1) the project was launched 10 years ago; (2) the developer mailing list of the project has archived over 1000 posts; and (3) the mailing list has been used by over 50 developers. The first criterion is to make sure that the project is non-trivial and has a long development period, the second criterion is to guarantee that there is rich data to extract architectural information from the mailing list of the project, and the last criterion is to select a project with a decent number of developers who collaborate and communicate in development. Since the goal of this study is to understand architecture information communication in OSS development spanning the life cycle of projects, each project must satisfy the above criteria even though a project may no longer be active now.

ArgoUML and Hibernate were finally selected according to the selection criteria. ArgoUML is an open source modeling tool, which is popular in the UML community. The developer mailing list of ArgoUML archived 26,647 posts from Jan 2000 to Aug 2020 with its releases evolving from v0.08 to v0.35.1. Hibernate implements the Java Persistence API which maps an object-oriented model (i.e., Java classes) to a relational database (database tables) and is widely used as the database middleware for Java applications. The developer mailing list of Hibernate accumulated 22,888 posts from Jan 2002 to Jul 2020 with its versions evolving from v0.9.1 to v5.4.19. The developer mailing list of Hibernate was not archived in one source, but was moved from SourceForge to JBoss partially due to the reason that a core developer discontinued his administration effort after Aug 2006[1]. The sources and durations of the developer mailing lists of ArgoUML and Hibernate investigated in this study are shown in Table 2. We checked the other communication channels of these two

---
[1] http://sourceforge.net/p/hibernate/mailman/message/13328372/

OSS projects that developers used. The developers of ArgoUML[2] used mailing lists and Wiki to communicate development knowledge, but the Wiki link is not accessible to us. The developers of Hibernate[3] had mailing lists, Wikis, and discussion forum for communication, but the Wikis and discussion forum are empty. As such, we chose mailing lists as the source to investigate and understand architecture information communication in these two OSS projects.

Table 2. ArgoUML and Hibernate developer mailing lists[4]

| OSS Project | URL of Developer Mailing List | Time Period | Num. of Posts | Num. of Architectural Posts |
|---|---|---|---|---|
| ArgoUML | http://argouml.tigris.org/ds/viewForumSummary.do?dsForumId=450 | Jan 2000 - July 2020 | 26,647 | 316 |
| Hibernate | http://sourceforge.net/p/hibernate/mailman/hibernate-devel/ | Jan 2002 - July 2006 | 8,913 | 257 |
| | http://lists.jboss.org/pipermail/hibernate-dev/ | Aug 2006 - July 2020 | 13,975 | |

## 3.4 Data collection

*1) Dataset*

In this study, we identify and collect architectural posts and architectural threads, a sequence of architectural posts on the same conversation topic, from all the posts in AgroUML and Hibernate developer mailing lists.

*2) Data collection method*

We collect all the posts in the mailing lists and manually identify architectural posts and threads from the collected posts of the two OSS projects selected in Section 3.3. To partially reduce the personal bias in architectural posts identification, we first performed a pilot identification of 30 posts with three researchers (i.e., the first three authors) by following the ISO 42010:2011 standard and any disagreement on the identification results was discussed and resolved by the four authors. We used ISO 42010:2011 to define architectural information in this study because it is the industry standard. ISO 42010:2011 describes a set of architecture elements that make up the architectural description model [25]. These architecture elements were used to identify and categorize architecture elements in an SA document in our recent survey on OSS SA documentation [16]. This categorization of architecture elements (listed in Table 3) is also used in this work to manually label and identify architecture information communicated in developer mailing lists (D2 in Table 4) because mailing lists are text-based which is similar to SA documents. The architecture elements with their definitions and examples identified from the mailing lists of the two OSS projects (see Section 3.3) are detailed in Table 3. After reaching a consensus on architectural posts, threads, and architecture elements identification through the pilot data collection, two reseachers collected data (i.e., architectural posts and threads) from the mailing lists of the two projects. The detailed information of data collection procedure is depicted below. In our analysis, we did not find any explicit discussions related to Architecture View, Architecture Viewpoint, Correspondence Rule, Correspondence, and Model Kind elements, and the potential reasons are discussed in Section 5.1.

---

[2] https://argouml-tigris-org.github.io/tigris/argouml/
[3] https://sourceforge.net/projects/hibernate/
[4] Note that, some provided links could be not working, as we noticed that the ArgoUML mailing list was down after 2020-07-01, but we provided all the collected architectural posts from the ArgoUML and Hibernate developer mailing lists online [47].

Table 3. Descrption of architecture elements in the ISO 42010:2011 standard [25] and their examples from the developer mailing lists

| Elements | Description | Example |
|---|---|---|
| **System of Interest** | refers to the system whose architecture is under consideration in the preparation of an architecture. | *High-level architecture of code generator in Hibernate*<br>Architectural post URL:<br>https://sourceforge.net/p/hibernate/mailman/message/5496885/ |
| **Stakeholder** | is an individual, team, or organization that has interests in a system. | *PS and GK (developers) who raised the concerns in the architectural post*<br>Architectural post URL:<br>https://sourceforge.net/p/hibernate/mailman/message/5496403/ |
| **Concern** | denotes an interest which pertains to the system's development, its operation or any other aspects that are critical or otherwise important to one or more stakeholders. Concerns include system considerations such as performance, reliability, security, distribution, and evolvability. | *My work on the XMI to HTML transform was to bring the existing transform into compliance with the XSLT 1.0 Recommendation.*<br>Architectural post URL:<br>http://argouml.tigris.org/ds/viewMessage.do?dsForumId=450&dsMessageId=116162 |
| **Architecture Model** | is developed using the methods established by its associated architectural viewpoint. | *There are four components using GEF: the Model component, the GUI Framework, the Diagrams and Internationalization.*<br>http://argouml.tigris.org/ds/viewMessage.do?dsForumId=450&viewType=browseAll&dsMessageId=223431 |
| **Architecture Rationale** | records explanation, justification, or reasoning about architectural design decisions that have been made in architecture design. | *One of the advantages is that you can choose which classes are initialized lazily at deployment time, simply by choosing which proxies to deploy.*<br>Architectural post URL:<br>https://sourceforge.net/p/hibernate/mailman/message/5496403/ |
| **Architecture View** | is a representation of a whole system from the perspective of a related set of concerns. | N/A |
| **Architecture Viewpoint** | is a specification of the conventions for constructing and using a view. | N/A |
| **Correspondence Rule** | is used to express, record, enforce and analyze consistency between models, views and other elements. | N/A |
| **Correspondence** | defines relations between above architecture elements, and correspondences are used to express consistency, traceability, dependencies, obligations, or other types of relations pertaining to the architecture being expressed. | N/A |
| **Model Kind** | is a convention for one type of modelling, such like data flow diagrams. | N/A |

*3) Data collection procedure*

To collect the architectural posts and threads in ArgoUML and Hibernate developer mailing lists, three steps are executed for data collection as detailed below (after the pilot identification mentioned above):

***Step 1*: Identification of architectural posts.** All the posts in the developer mailing lists of ArgoUML and Hibernate are manually and independently checked by two researchers (i.e., the first and second authors), and any posts which the two researcher have different opinions on their types were further discussed and confirmed with another researcher (i.e., the third author).

Furthemore, architectural posts are identified and classified according to the architecure elements (Table 3) defined in the ISO 42010:2011 standard [25]. If a post contains only System of Interest, it is not considered as an architectural post, because this element alone does not provide sufficient architectural information. The architecture elements in the architectural posts were identified by the first and second authors, and further examined by the third and fourth authors, to mitigate personal bias in the identification. In addition, we made a final reliability test on architectural posts identification, and calculated Cohen's Kappa reliability coefficient [46] for the identification between the two annotators, which is 0.85.

***Step 2*: Identification of architectural threads**. One architectural thread is composed of a sequence of (one or many) architectural posts on the same conversation topic. Each architectural post is manually read and checked by two authors (i.e., the first and second authors) independently to decide the architecture thread which it belongs to, and any controversial architectural threads were further discussed with the third author.

***Step 3*: Related data collection**. Data items (i.e., D1-D6 described in Table 4) about architectural posts and threads for data analysis (see Section 3.5) are collected and recorded in an Excel spreadsheet (which has been made available online [47]).

Table 4 shows the mappings from the extracted data items of OSS developer mailing lists to the four RQs. The purposes of architecture information communication using mailing lists (RQ1) can be directly answered using the data of D4. The role of developers (D5) and their related architecture information communication activity (D3) are used to analyze how architectural information is communicated between roles in OSS projects using mailing lists (RQ2). By combining post time and release information (D1 and D6) with communication activity (D3), we can answer when OSS developers tend to communicate architecture information using mailing lists (RQ3). D2 can be directly used to answer what types of architectural information are communicated using mailing lists (RQ4).

We opted to analyze the architectural posts manually. We did not employ a (semi-)automated approach in this study to identify architectural posts, which may miss or mistakenly classify some of the architectural posts and consequently threaten the validity of the results.

Table 4. Extracted data items and their mappings to research questions (RQs)

| # | Data Item | Description | RQ |
|---|-----------|-------------|-----|
| D1 | Post time | Record the time when a post that contains architectural information was posted. | RQ3 |
| D2 | Architecture information elements communicated | Record the type and content of architecture information elements in the architectural posts. We used the criteria employed in our recent survey on OSS architecture documentation [16], which are elaborated in Section 3.4, to judge whether or not a post contains architectural information. | RQ4 |
| D3 | Architecture information communication activity | Denote the type of architecture information communication activities of the architectural posts. We used three predefined communication activities (*ask*, *mention*, and *reply*), to specify how architecture information is communicated in mailing lists. When an architectral post is used to inquire or answer an architecture issue/question, we name this activity *ask* or *reply* activity. Other communications are regarded as *mention*. | RQ2 RQ3 |
| D4 | Purpose of architecture information communication | Record the aim of architecture information communication of the architectural posts. We used a bottom-up encoding approach to identify and classify the purposes of architecture information communication, which is elaborated in Section 4.1. | RQ1 |
| D5 | Role of developer | Denote the roles (*initiator*, *asker*, and *responder*) of developers who communicate architecture information in an architectural thread using mailing lists. Developers who post the first post in an architectural thread play the role of *initiator*, who may conduct the communication activities (D3) *ask* or *mention*; developers who raise an architecture issue/question are regarded as *asker*, who conduct the communication activity *ask*; and developers who respond an architectural post is a *responder*, who conduct the communication activity *reply*. | RQ2 |
| D6 | Release version and time | Record the version and time of the release which is immediately after the date of an architectural post. This data term can be used with D1 (post time) to analyze the relationship between releases and frequency of architecture information communication in development in a time perspective. | RQ3 |

## 3.5 Data analysis

The extracted qualitative data (D2, D3, D4, D5) and quantitative data (D1, D6) from the architectural posts and threads are analyzed to answer the RQs defined in Section 3.1.

Bottom-up approaches are suitable for classifying specific domain knowledge as concepts when there is no predefined and existing concepts in that domain [28]. We employed a bottom-up and systematic encoding and analysis approach from Grounded Theory [1] to manually analyze the qualitative data (i.e., the purposes of architecture information communication "D4" for answering RQ1 and the detailed content of architecture information communication "D2" for answering RQ4). In our analysis, only open coding and axial coding techniques from Grounded Theory were employed. The process comprises three steps which were executed iteratively [1]: (1) open coding, executed by the first two authors, split the communicated information in architectural posts (i.e., data) into separate parts, i.e., words, phrases, or sentences, which are labelled as concepts (e.g., developers of ArgoUML discussed the user interface (UI) design of the navigator[5], which is named as a concept "UI adjustment"), (2) axial coding, executed by the first two authors and confirmed with the third author, was then employed to identify categories (i.e., main topics and subtopics) through refining and relating the concepts generated in open coding to a category (e.g., concept "UI adjustment" is regarded as a type of adaptation when the environment changes, and is related to a core category "Adaptive environment"), and (3) to reduce the personal bias in coding, any inconsistencies on the coding results are cross examined by the all the authors. For example, for answering RQ4 (i.e., what architecture information is communicated in OSS development), we summarized and coded 4 main topics and 23 subtopics from the identified architectural posts for investigating and structuring what architecture information developers discussed. As one developer posted "*I am part way through the job of refactoring Hibernate so that SQL generation occurs in its own layer*", and we coded "Refactoring" as the subtopic of this post that was further classified into "Architecture" topic (see Table 5). All the data (i.e., the architectural posts and elements of the two OSS projects) and coding results of architecture information have been made available online [47].

---
[5] http://argouml.tigris.org/ds/viewMessage.do?dsForumId=450&dsMessageId=116641

Descriptive statistics are used to analyze the quantitative data (D1 and D6) in OSS developer mailing lists for answering RQ3 (architectural posts and threads publishing frequency). We get the quantitative data (number of communications) based on the qualitative data (D3 and D5) in OSS developer mailing lists for answering RQ2 (social network of architecture information communication).

## 4. Results

The results of the case study are elaborated in this section. We describe the purposes of architecture information communication using mailing lists (answer to RQ1) in Section 4.1. The social networks of architecture information communication (answer to RQ2) are presented in Section 4.2. The frequency and timing of architectural posts and threads (answer to RQ3) in a (release) time perspective is provided in Section 4.3. The content of architecture information communication in mailing lists (answer to RQ4) is elaborated in Section 4.4.

### 4.1 Purposes of architecture information communication (Why)

We analyze and identify the purposes of architecture information communication through encoding. Encoding of posts has gone through an iterative process. For example, in a post one developer was against the idea of Hibernate Search DSL proposed by another developer[6]. The first and second authors initially coded this post as architecture objection in open coding. After a discussion between all the authors during selective coding, we decided that an objection with reasons is part of the process of negotiation in architecture design, and classified such posts as architecture negotiation. Fig. 1 and Fig. 2 present the distribution of numbers of posts of the purposes of architecture information communication in ArgoUML and Hibernate, respectively.

We divided the studied period of the two projects into two stages according to their first stable releases, i.e., ArgoUML v0.10[7] and Hibernate v1.0.0 final[8], to investigate the relationship between timing of version release and purposes of architecture information communication, i.e., how architecture information is communicated before and after the first stable version is released. The purposes of architecture information communication identified from the architectural posts in this study are the following:

---

[6] http://lists.jboss.org/pipermail/hibernate-dev/2009-November/004472.html
[7] http://argouml.tigris.org/servlets/NewsItemView?newsItemID=128
[8] http://sourceforge.net/p/hibernate/mailman/message/5497028/

**Architecture suggestion**: This architecture information communicated is used to guide the work of other developers, e.g., a developer suggested new features for the next version of Hibernate project with related design decisions[9].

**Architecture update notification** is used to announce the update or modifications of architecture, e.g., a developer announced that he made a new release for Eclipse GEF v0.9.3 and described how to use a framework developed by Gentleware to annotate modeling elements[10].

**Architecture negotiation**: The architecture issues and questions, e.g., about adding new components, are discussed between developers. For example, a developer proposed an architecture solution (decoupling the graphical objects from the model elements to follow the MVC pattern[11]), and negotiated with other developers this solution/modification. The difference between architecture negotiation and suggestion is that the suggested solutions are just followed by other developers without any objection and negotiation.

**Architecture interpretation**: This architecture information is used to clarify the understanding of architecture, e.g., a developer explained the differences of architecture between Hibernate and Castor[12].

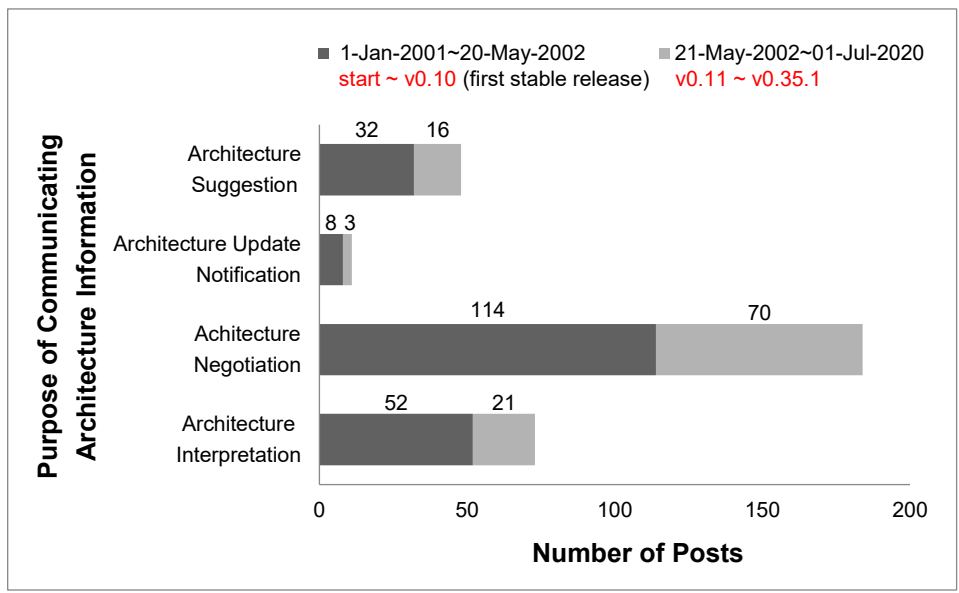

Fig. 1. Distribution of numbers of architectural posts over purposes of architecture information communication in the mailing list of ArgoUML

---

[9] http://sourceforge.net/p/hibernate/mailman/message/5496381/
[10] http://argouml.tigris.org/ds/viewMessage.do?dsForumId=450&dsMessageId=132373
[11] http://argouml.tigris.org/ds/viewMessage.do?dsForumId=450&dsMessageId=130478
[12] http://sourceforge.net/p/hibernate/mailman/message/5496425/

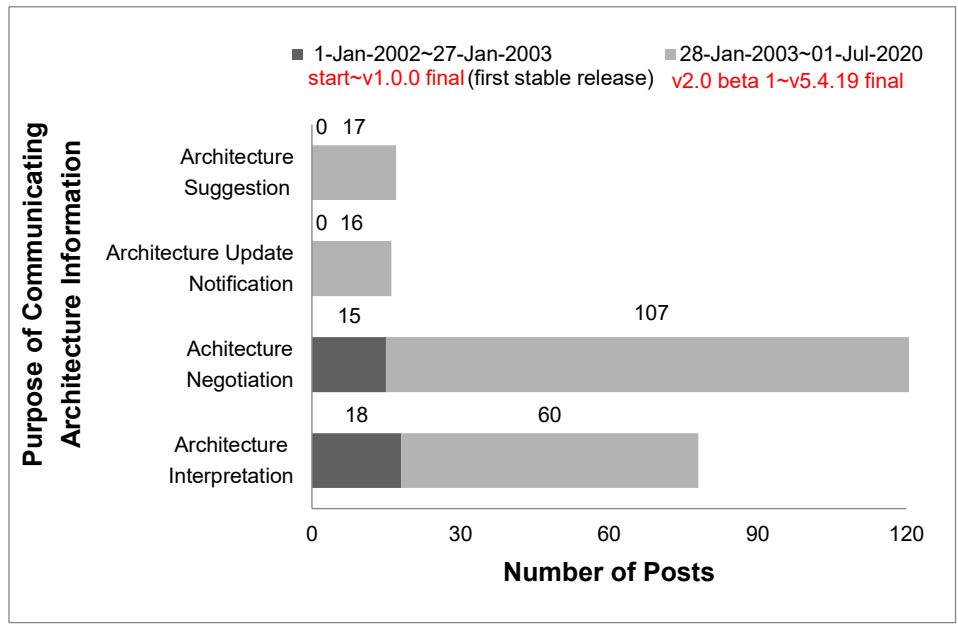

Fig. 2. Distribution of numbers of architectural posts over purpose of architecture information communication in the mailing list of Hibernate

**RQ1 Findings**: As shown in Fig. 1 and Fig. 2, architecture negotiation and architecture interpretation are the main purposes of architecture information communication through developer mailing lists in both OSS projects. The purposes of architecture information communication varied in different development phases, e.g., developers in ArgoUML mainly communicated architecture information to negotiate and explain architecture issues at the early stage of development (Fig. 1).

## 4.2 Social network of architecture information communication (Who)

One key factor to the success of OSS projects is that all the activities are conducted in public, and mailing lists can trace the communication and coordination of the participants in OSS development [9]. Dietze found that core developers modified the majority of source code and a close interaction took place between core developers and maintainers in OSS development [14]. However, core developers can be identified from different perspectives (e.g., coding, architecting), and developers who communicated information in mailing lists may not be the ones who modified source code, e.g., there are 79 developers who communicated architecture information in the ArgoUML developer mailing list and there are only 45 developers who participated in the ArgoUML commit mailing list (we regard participants who communicated information in developer mailing lists as developers or potential developers of the OSS projects). In this work, we adapt and follow the identification method of core developers proposed in [32]: *those that were actively involved in higher levels of knowledge sharing* (i.e., architecture information communication in this work). To identify the structure of social network and the core developers (nodes) of architecture information communication in OSS development through mailing lists, we employ two network measures, which are fundamental in describing developer social networks in software development and meaningful in the context of architecture information communication: degree centrality and betweenness centrality [45]. *Degree centrality* denotes the number of nodes that directly connect to a node [19]. Developers with high degree centrality are considered the central point of architecture information communication and the major channel of architecture knowledge sharing. For example, developers tend to directly communicate with the developer who has the most knowledge about the architecture of the system. *Betweenness centrality* measures the number of communications a node acts as a bridge along the shortest path between two other nodes [20]. Hence, two separate developers can communicate their architecture information through the developers with high betweenness centrality. For example, a core developer of a subsystem may act as a bridge between the developers of the subsystem and the architect of the system when s/he could not answer certain questions about the architecture of the system.

To identify the core developers in the social networks of architecture information communication, we checked the developers with higher scores using the two centrality measurements (i.e., the top ten scores by summing the normalized degree centrality and betweenness centrality, which are transformed to a value between 0.0 and 1.0 using min-max normalization). The results of the social networks of architecture information communication in the mailing lists of ArgoUML and Hibernate are illustrated in Fig. 3 and Fig. 4 respectively, which are calculated and visualized by the Gephi tool[13]. We use grey and black circles to denote the core and general developers respectively. Note that the ALL circle in white is not a developer, but represents the posts sent to all the members of the mailing lists (i.e., without sending to one or several specific receivers), and this node was excluded when we calculated the two centrality measurements using Gephi.

We use data item D3 (architecture information communication activity) and D5 (role of developer) in Table 4 to construct the structure of the social networks of architecture information communication. If an *initiator/asker* only *mentions/asks* architecture information without any *responders*, we identify that the developer is broadcasting to all the members of the mailing lists and connect her/him to the circle ALL. If an architectural post from a developer is used to *ask* an architecture issue to or *reply* an architecture issue from another developer, we directly connect those two developers. Note that we read the posts instead of only looking at the header content of the posts to identify the communication (e.g., *ask* or *reply*) between developers, because not all the response posts in one architecture thread start with "*Re:*" or the same title in the header. Initials in circles are used to denote core developers, e.g., AC[14]. The number beside a line shows the number of architectural threads communicated between two developers, e.g., AC and TS communicated architecture information in five threads. The line without a number means that the two developers communicated only in one thread. From Fig. 3, we find that architecture information communication happens intensively amongst several core developers (e.g., AC and BT) in ArgoUML. From Fig. 4, we see a similar social network analysis results in Hibernate, and GK and EB are the core developers of Hibernate. We find that the developers of ArgoUML broadcasted architectural information more frequently than the developers in Hibernate (39.5% vs. 18.3%). We also find some core developers whose centrality scores are lower than GK, but higher than most other developers, e.g., MRA in Fig. 4. These developers act as the core developers in second-tier.

---

[13] http://gephi.org/
[14] Only the abbreviations of the developers' names are provided due to privacy concerns.

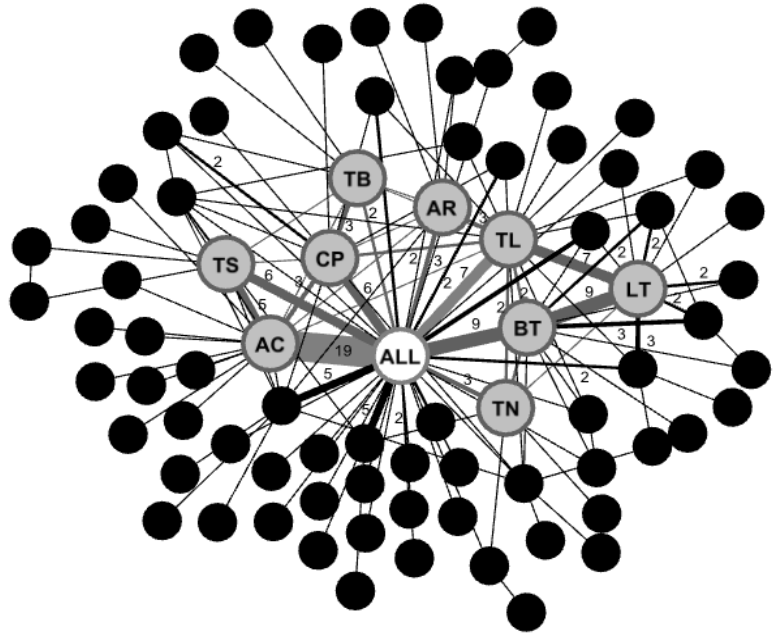

Fig. 3. Social network graph of communicating architecture information between developers in ArgoUML

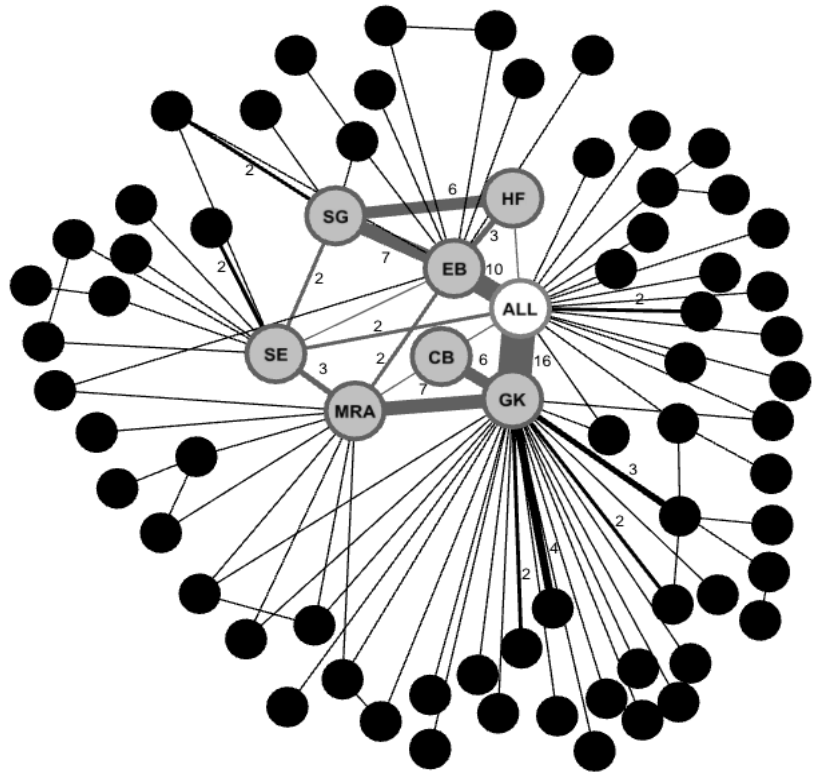

Fig. 4. Social network graph of communicating architecture information between developers in Hibernate

**RQ2 Findings**: There are not many developers who are involved in architecture communication. According to the results shown in Fig. 3 and Fig. 4, there are nine core developers in ArgoUML and seven core developers in Hibernate. In these two projects, architecture information communication activities happened intensively amongst a few developers, e.g., 79.0% architecture information communication happened amongst nine core developers in ArgoUML (158 out of 200 threads); and 69.2% amongst seven core developers in Hibernate (110 out of 159 threads). These results are similar to the findings reported in random networks [6] that "*networks expand continuously by the addition of new nodes, and new nodes attach preferentially to already well-connected sites*". As shown in Fig. 3 and Fig. 4, the new participants tend to communicate with these core developers. The results are also in line with the finding reported in a study on the central role of mailing lists in OSS development [40] that "*mailing list activity is driven by a dominant group of participants*".

## 4.3 Architectural posts and threads publishing timing (When)

We use histograms to depict the numbers of architectural posts and threads that were published per month during the investigated time period in the two OSS projects, and marked the release versions and dates of the OSS in the histograms. The results in Fig. 5 and Fig. 6 show that the number of posts and threads that communicated architecture information had decreased during the investigated period in both OSS projects. We observe peaks in the number of architectural posts and threads per month before the first stable release dates, e.g., first release and release candidate (RC). For example, there are 28 architectural posts and 17 architectural threads published during the five months before Hibernate v1.0.0 final was released, and the numbers decreased to 12 posts and 10 threads during the five months after Hibernate v1.0.0 final was released.

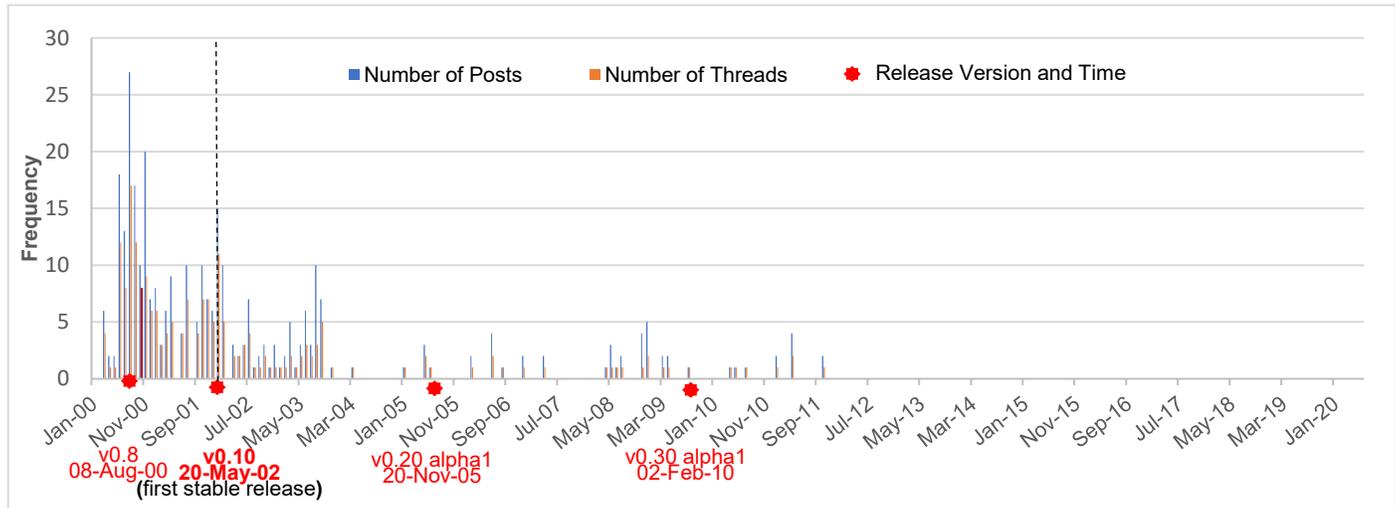

Fig. 5. Numbers of architectural posts/threads and releases in ArgoUML

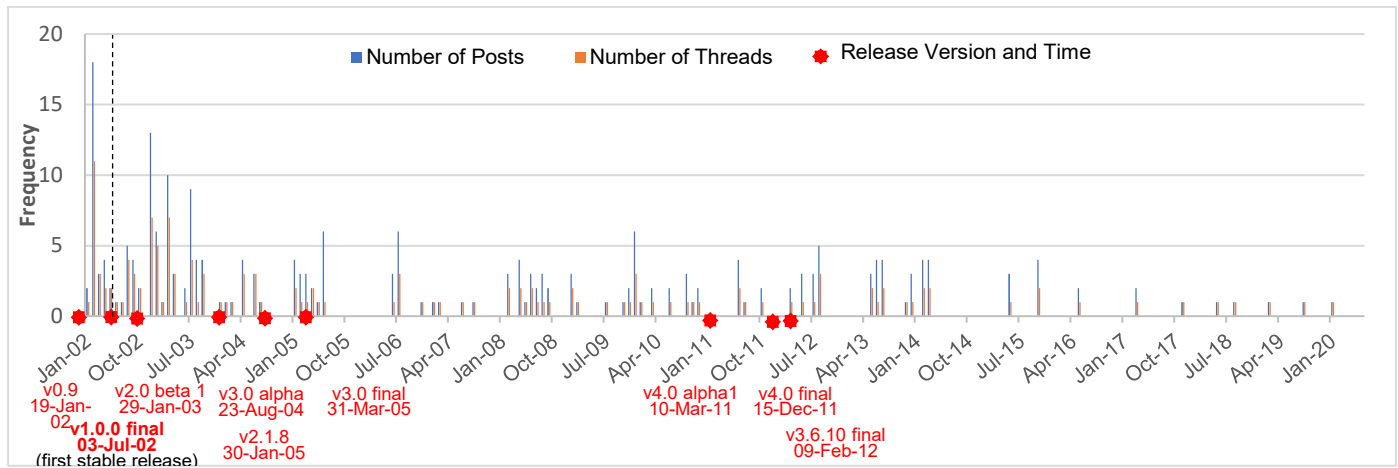

Fig. 6. Numbers of architectural posts/threads and releases in Hibernate

**RQ3 Findings**: The frequencies of architecture information communication are different between ArgoUML and Hibernate. In ArgoUML, architecture information was largely communicated during the early stage of OSS development, i.e., the phase when an OSS project is prepared to release the first stable version. For example, 71% (142 out of 200) architectural threads and 65.2% (206 out of 316) architectural posts were communicated before ArgoUML v0.10 was released on 20-May-2002. In Hibernate, 11.9% (19 out of 159) architectural threads and 11.6% (30 out of 257) architectural posts were communicated before Hibernate v1.0.0 final was released on 03-July-2002. One common result between ArgoUML and Hibernate is that the number of architectural threads that communicate architecture information through mailing lists decreases (for over 18 years in both projects) after the first stable releases, as shown in Fig. 5 and Fig. 6.

## 4.4 Architecture information communicated during development (What)

The conceptual model for architectural description in the ISO 42010:2011 standard [25] is used to identify the communicated architectural information in mailing lists. Fig. 7 provides the distribution (in percentage) of architectural posts over architecture elements communicated in the developer mailing lists of ArgoUML and Hibernate. The distribution of three communicated architecture elements Architecture Rationale, Architecture Model, and System of Interest in both projects is almost the same. The sum of the percentages of architectural posts in Fig. 7 is greater than 100%, because one post can communicate several architecture information elements (e.g., System of Interest and Concern).

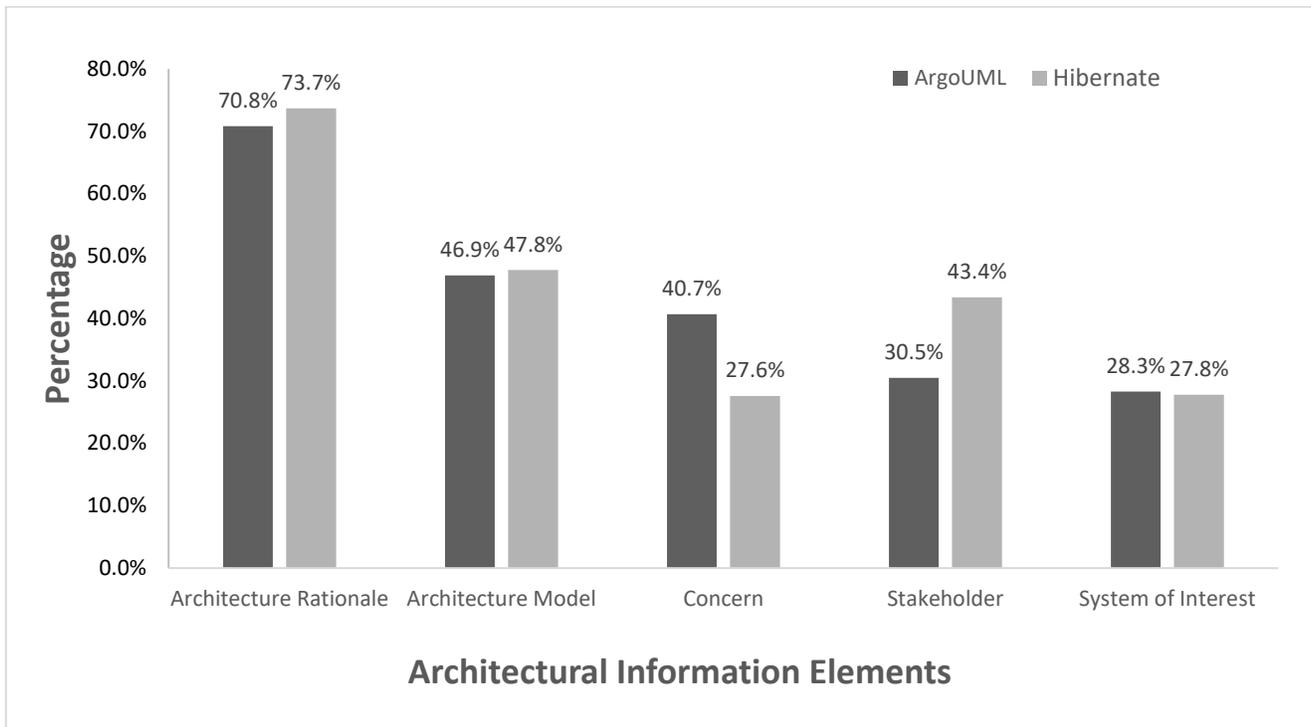

Fig. 7. Distribution of architectural posts over architecture elements in the mailing lists of ArgoUML and Hibernate

We analyzed the detailed contents of architecture information communication (as described in Section 3.5) and coded the content of architecture information communication with 23 subtopics. Requirements and architecture were discussed at different abstraction levels, and consequently we classified the 23 discussion subtopics into four categories: **architecture topic** (i.e., specific design topics about the architecture), **abstraction level** (i.e., the level of the architecture artifacts discussed in architecture information communication), **external quality requirement** (i.e., the quality characteristics of a system from the perspective of its users satisfied by architecture changes), and **internal quality requirement** (i.e., the quality characteristics of a system from the perspective of its developers satisfied by architecture changes). Table 5 provides the detailed descriptions and examples of the 4 categories and 23 subtopics, which are an extension of the codes identified in our previous work that specifically focuses on the causes of architecture changes communicated in mailing lists [15]. The percentages of each subtopic (i.e., content of architecture information communication) communicated in architectural threads before and after the first stable release (BFR and AFR) in ArgoUML and Hibernate are shown in Fig. 8 and Fig. 9 respectively. Note that architectural thread is used here as the unit to calculate the percentages, because architectural thread is a logical unit of conversation and it makes sense to count the number of occurrences of a code (e.g., Modularity) in an architectural thread only once even if developers discuss the code multiple times in the thread.

**Table 5.** Coding results of the content of architecture information communication in mailing lists

| Category | Subtopic Code | Description | Example Post (with URL to Access Post) |
|---|---|---|---|
| **Architecture Topic** | New component | Adding a component in the system to satisfy a functional or system requirement. | *I would prefer to create a Persistence Manager Component that uses Hibernate underneath.*<br>Architectural thread URL:<br>https://sourceforge.net/p/hibernate/mailman/message/11229331/ |
| | New feature | Adding a new feature to the system. | *Adding a new feature of command line or GUI toolset for (a) generation and execution of table schema (b) automated generation of skeletal XML mappings.*<br>Architectural thread URL:<br>https://sourceforge.net/p/hibernate/mailman/message/5496381/ |
| | Refactoring | Discussion about adapting existing architecture for preventing the prone to architectural defects. | *Two separate associations were used to replace one-to-many association in the Hibernate database model, which makes the architecture less error-prone.*<br>Architectural thread URL:<br>https://sourceforge.net/p/hibernate/mailman/message/5496466/ |
| | Use of design principle | Employing design rules, approaches, or patterns to meet design requirements. | *Think about the Observer Pattern for this. So we need an Observer which listens to state changes in the Controller/View of the Presentation Layer.*<br>Architectural thread URL:<br>https://sourceforge.net/p/hibernate/mailman/message/7780586/ |
| | Adaptive environment | Architecture adjustment for adapting to a new environment, platform, or standard. | *It's time to think about building on top of that for the long overdue migration from UML 1.4 to the actual UML 2.x version.*<br>Architectural thread URL:<br>http://argouml.tigris.org/ds/viewMessage.do?dsForumId=450&viewType=browseAll&dsMessageId=1492835 |
| | Component dependency | Discussion about how a component depends on other components. | *Core module depends on envers module when adding envers event in Hibernate.*<br>Architectural thread URL:<br>http://lists.jboss.org/pipermail/hibernate-dev/2011-June/006630.html |
| | Alternative | Providing alternative solutions for an existing architecture design. | *To integrate with transactions in Hibernate, developer provide alternative approaches to make use of Hibernate and acme integration.*<br>Architectural thread URL:<br>http://lists.jboss.org/pipermail/hibernate-dev/2012-August/008920.html |
| **Abstraction Level** | System level | Architecture information is discussed at the system level about system or subsystem's architecture. | *Integration between the search engine (Hibernate Search) and Hibernate ORM without losing the search engine power.*<br>Architectural thread URL:<br>https://sourceforge.net/p/hibernate/mailman/message/13328113/ |
| | Package level | Architecture information is discussed at the package level. | *Refactoring org.hibernate.tuple package.*<br>Architectural thread URL:<br>https://sourceforge.net/p/hibernate/mailman/message/13328213/ |
| | Class level | Architecture information is discussed at the class level. | *Duplicating several utility classes into hibernate-search-testing module in Hibernate.*<br>Architectural thread URL:<br>http://lists.jboss.org/pipermail/hibernate-dev/2010-March/004975.html |
| **External Quality Requirement** | Customization | Meeting to customized requirements from various users. | *The design of the code generators in ArgoUML allows users to write their own tools for generating code.*<br>Architectural thread URL:<br>http://argouml.tigris.org/ds/viewMessage.do?dsForumId=450&dsMessageId=131107 |
| | Efficiency | Improving the performance of the system with available resources. | *Achieving the required response time with limited memory.*<br>Architectural thread URL:<br>https://sourceforge.net/p/hibernate/mailman/message/11229735/ |
| | Extensibility | The system can be extended to accommodate new environments. | *The query language has the potential to be one of the best features of the project and I can think of a few really great extensions.*<br>Architectural thread URL:<br>https://sourceforge.net/p/hibernate/mailman/message/5496381/ |
| | Localization | Adapting the system to fit special countries/regions. | *Committed some new classes and patches to GEF that GEF to be localized, i.e., you can have an English, German, or French version.*<br>Architectural thread URL:<br>http://argouml.tigris.org/ds/viewMessage.do?dsForumId=450&dsMessageId=133436 |
| **Internal Quality Requirement** | Conciseness | Making the architecture elements (e.g., classes/packages/components) simple and readable (e.g., through architecture refactoring). | *Define ArgoUML's internal structure much more clearly.*<br>Architecture thread URL:<br>http://argouml.tigris.org/ds/viewMessage.do?dsForumId=450&dsMessageId=136637 |
| | Consistency | The architecture design elements should be consistent with each other. | *The wrapper itself never knows how fully initialized it is. This is consistent with the current design.*<br>Architectural thread URL:<br>https://sourceforge.net/p/hibernate/mailman/message/5496534/ |
| | Compatibility | New components added to the system should be compatible with existing components/systems/technologies. | *I committed into cvs changes to the Hibernate XDoclet module. These changes should bring the module into compliance with the hibernate-mapping-1.1.dtd.*<br>Architectural thread URL:<br>https://sourceforge.net/p/hibernate/mailman/message/7780402/ |
| | Completeness | The architecture design of the system can address all the requirements in the specification. | *Completion of the EntityMode and Tuplizer capabilities.*<br>Architecture thread URL:<br>http://sourceforge.net/p/hibernate/mailman/message/13328208/ |
| | Flexibility | The architecture of the system can be easily adapted to address new requirements. | *I thought how to achieve maximum flexibility and keep a clean code for the cache, and I came up with the following design that makes more sense.*<br>Architecture thread URL:<br>https://sourceforge.net/p/hibernate/mailman/message/5497820/ |
| | Modularity | The degree to which a system is composed of discrete components such that a change to one component has a minimal impact on other components. | *I was hoping to spend a few hours next week to see if I could decouple Argo from XML4J.*<br>Architecture thread URL:<br>http://argouml.tigris.org/ds/viewMessage.do?dsForumId=450&dsMessageId=116201 |
| | Traceability | The ability of tracing from one architecture element (e.g., class/package/component) to another. | *Logically split functionality into modules and define what is "ArgoUML: Basic Edition". Trace dependencies between ArgoUML base functionality and other parts.*<br>Architecture thread URL:<br>http://argouml.tigris.org/ds/viewMessage.do?dsForumId=450&viewType=browseAll&dsMessageId=116448 |
| | Understandability | The degree of how easy developers can understand the architecture design of the system. | *The event mechanism used in ArgoUML to synchronize model and graphical objects is spread over lots of classes and therefore hard to understand and error prone.*<br>Architecture thread URL:<br>http://argouml.tigris.org/ds/viewMessage.do?dsForumId=450&viewType=browseAll&dsMessageId=130478 |
| | Reusability | Reusing existing architecture in the following releases of a project. | *What to reuse. We will most likely use Jandex to read annotations to benefit from Hibernate ORM.*<br>Architecture thread URL:<br>http://lists.jboss.org/pipermail/hibernate-dev/2013-May/009802.html |

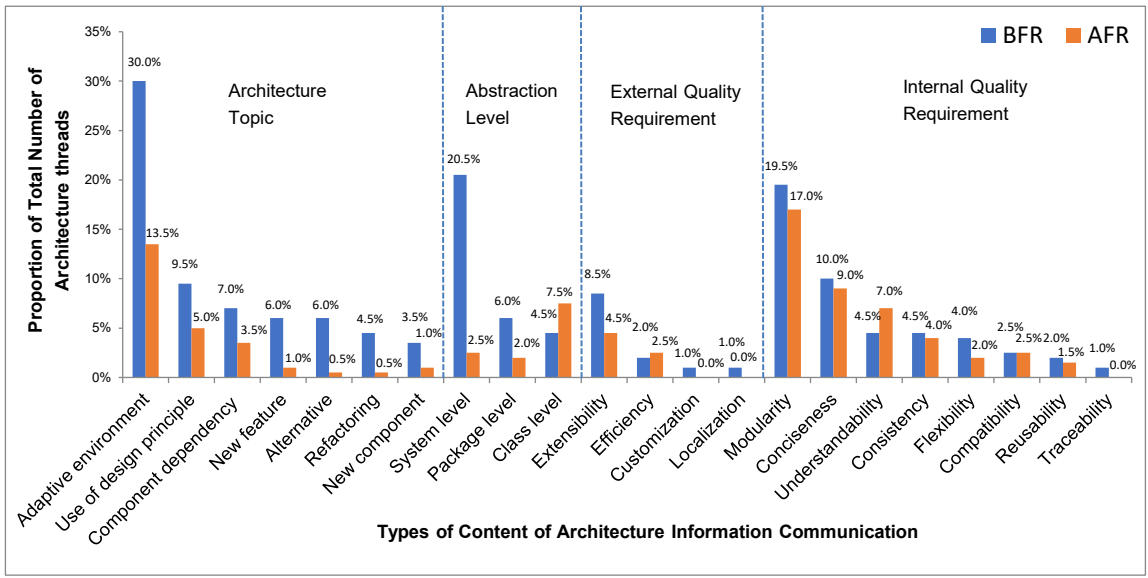

Fig. 8. Proportions of the content of architecture information communication in ArgoUML before and after the first stable release

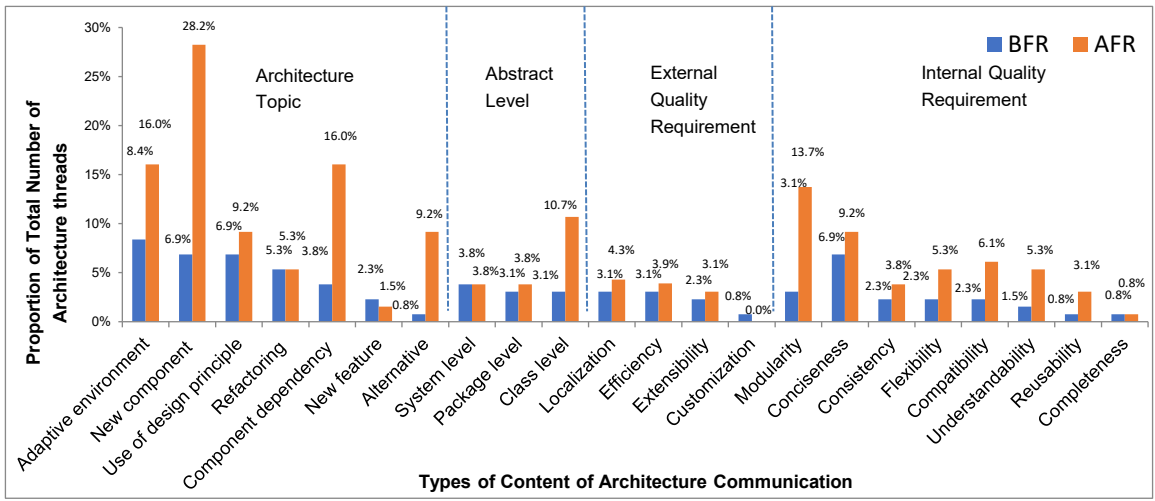

Fig. 9. Proportions of the content of architecture information communication in Hibernate before and after the first stable release

To investigate the interconnection between architecture topic and quality requirement discussion in the mailing lists, we counted the numbers of the co-occurrence of these two types of codes in architectural threads, which are shown as bubbles in Fig. 10 and Fig. 11 for ArgoUML and Hibernate, respectively. For example, the bubble with the number 29 in Fig. 10 means that "Adaptive environment" and "Modularity" were discussed together in 29 architectural threads in ArgoUML. When we investigated the architecture design topics (i.e., the nature of the software development) and the quality requirements (i.e., what is required of the development), we found that there are different characteristics of the two systems. ArgoUML team spent much of its effort in looking into "Adaptive environment" and "Component dependency". The numbers of threads were 75 and 67, respectively (see Fig. 10). The main quality requirements they discussed and focused on were "Modularity" and "Conciseness" as shown in Fig. 10. Hibernate team, on the other hand, posted across all architecture topics, and their discussions on "Modularity", "Conciseness", "Efficiency", "Consistency", and "Extensibility" were evenly spread (see Fig. 11). An example in ArgoUML, developers recommended to use generic controls with reflection mechanism in plain Java to make their code easy to maintain ("Conciseness") when employing the UML data model provided by NSUML ("Adaptive environment")[15]. An example in Hibernate, to add *new Dialects* in Hibernate, developers discussed whether to integrate a new class *org.hibernate.dialect.HiRDDialect* ("New component") into Hibernate or use an "*extras*" module that can encapsulate code ("Modularity")[16].

---

[15] http://argouml.tigris.org/ds/viewMessage.do?dsForumId=450&viewType=browseAll&dsMessageId=136836#messagefocus
[16] http://lists.jboss.org/pipermail/hibernate-dev/2008-September/003323.html

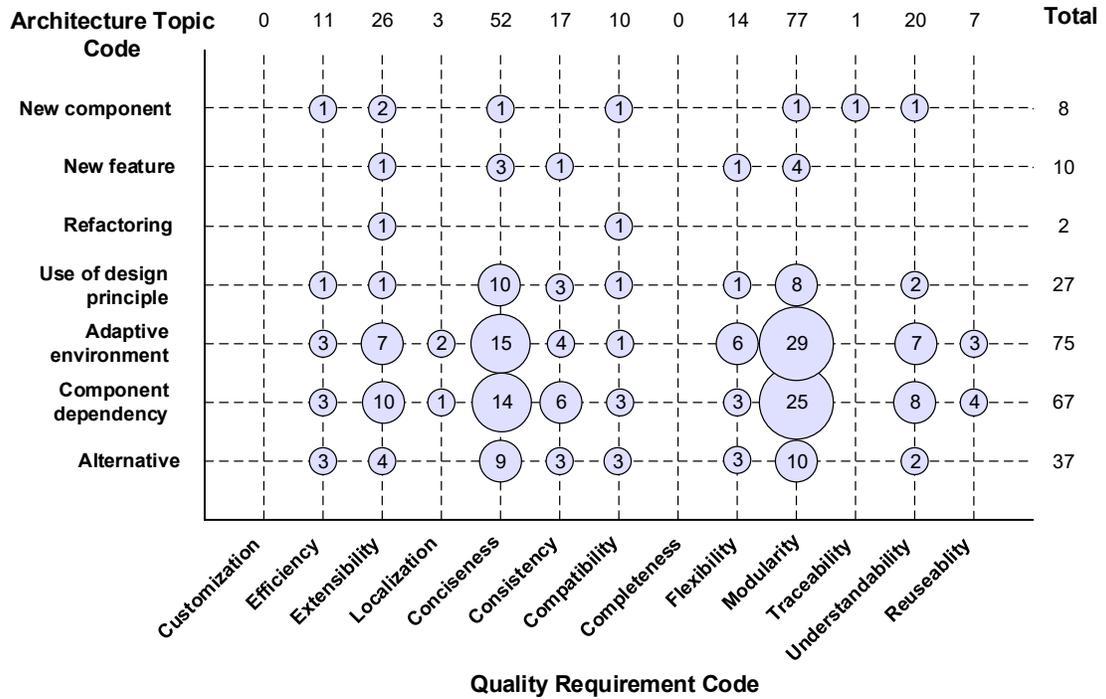

Fig. 10. Interconnection between architecture topic and quality requirement codes in the architectural threads of ArgoUML

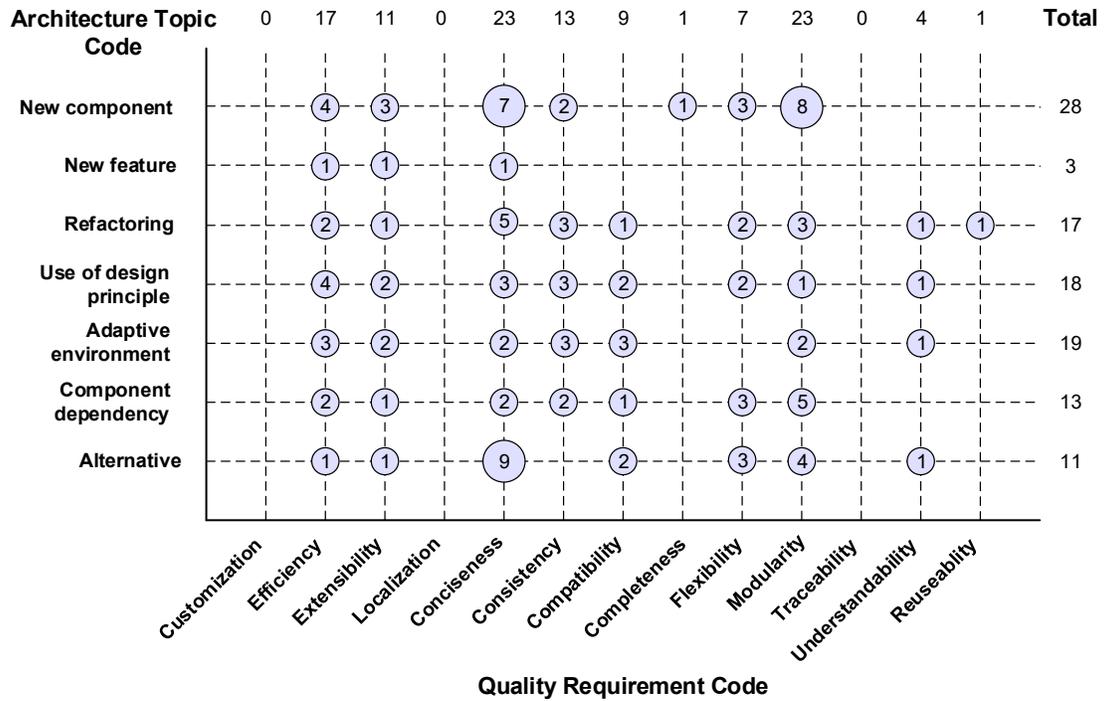

Fig. 11. Interconnection between architecture topic and quality requirement codes in the architectural threads of Hibernate

**RQ4 Findings**: The architecture elements that were most frequently communicated in the mailing lists of both projects are Architecture Rationale and Architecture Model. The frequently communicated content of architecture in ArgoUML and Hibernate is partially different. The frequently communicated content is "Adaptive environment" in ArgoUML and "New component" in Hibernate. The frequently communicated external quality requirement is "Extensibility" in ArgoUML and "Efficiency" in Hibernate. The frequently communicated internal quality requirement is "Modularity" in both projects. The amount of communicated content of various architecture information follows a similar trend before and after the first stable release in AgroUML (e.g., "Adaptive environment" dominates the discussion before and after the first stable release), which is not the case in Hibernate (e.g., "Class level" is the least communicated architecture content before the first stable release, but became the most frequently communicated content after the first stable release). Architecture topics and quality requirements were extensively discussed together in both projects with different characteristics, which is an evidence of relating requriements and architectures [3].

## 5. Discussion

In this section, we first interpret the results of the four RQs (i.e., why, who, when, and what) presented in Section 4. We then discuss the implications of the results with a summary of them.

### 5.1 Interpretation of results

> **Why**: Architecture negotiation and interpretation are the main purposes of architecture information communication, and these two aspects of architecture information communication help architects make decisions in OSS development.

**Architecture negotiation and interpretation**: From Fig. 1 and Fig. 2, we found that the proportion of architecture suggestion in all purposes of architecture information communication in ArgoUML, 15.2% (48 out of 316 architectural posts), is significantly higher than that in Hibernate, 7.3% (19 out of 257 architectural posts). One interpretation is that ArgoUML employed a high-level architecture pattern at the initial stage of development, i.e., the MVC (Model-View-Controller) pattern. The core developers provided their suggestions for using this architecture pattern, e.g., "*decouple the graphical objects from the model elements by introducing dedicated controller classes*", which follows the MVC pattern[17].

---
[17] http://argouml.tigris.org/ds/viewMessage.do?dsForumId=450&dsMessageId=130478

Architecture negotiation and interpretation are the two main purposes of architecture information communication in both ArgoUML and Hibernate (58.2%, 23.1% and 55.5%, 34.3% respectively). Most of the negotiations and interpretations of Hibernate were carried out after the first stable release, whilst ArgoUML had most of the negotiations and interpretations before the first stable release. Despite the timing of architecture negotiation and interpretation between the two OSS projects, developers of both projects carried out signification amount of negotiation (i.e., negotiating a trade-off between conflicting challenges) and interpretation (i.e., clarifying architecture design). The negotiations and interpretations correlate with adapting the environment in ArgoUML and the new component development in Hibernate (see Fig. 8 and Fig. 9). Another reason for architecture negotiation and interpretation is for decision making in architecture design. Architects communicate to identify potential mismatches between conceptual design and actual implementation of the architecture. Kruchten [30] suggests that architects communicate frequently. We have provided empirical evidence to support that this is indeed the case, and further found that architects spend most of the communication in negotiating and interpreting architectural issues, especially around the time of releases. This finding puts into focus the necessity to *educate architects to carry out architecture interpretation and negotiation* as part of their professional practice.

> **Who**: Architecture information communication took place intensively amongst few core developers, and core developers changed during OSS development.

**Architecture information communication mainly between few core developers**: Architecture information is negotiated between few core developers of an OSS project who make architectural decisions, instead of extensively discussing the decisions among all developers which is time-consuming. For example, from Fig. 3 and Fig. 4, we found that most architectural threads have *responders*, and there are 86.8% (138 out of 159) architectural threads communicated between developers in Hibernate, i.e., they were not broadcasting their architectural information to all the members of the mailing list. The architectural threads were mainly communicated between two or three developers. Furthermore, there are only 9.4% (15 out of 159) architectural threads that have more than three participants in Hibernate. As such, being aware of those core developers could help the development community to *identify decision makers* ("*who knows what*"), which can facilitate the implementation of an effective personalization strategy in architectural knowledge management [50].

**Change of core developers**: The core developers who communicated architectural information varied in different stages of OSS development in both ArgoUML and Hibernate. For example, GK participated in 68.3% architectural threads (28 out of 41) before Hibernate v2.0 beta1 was released on 29-Jan-2003, while GK was only involved in 10.2% architectural threads (6 out of 59) after Hibernate v3.0 alpha was released on 23-Aug-2004. EB dominated the communication of architectural threads after Hibernate evolved to v3.0, i.e., he participated in 40.7% architectural threads (24 out of 59). This result is similar to the result reported in [44], which indicates that core developers can be different people in different release versions. This shows that different developers make architectural design decisions in different versions.

> **When**: There is a relationship between the timing of architecture information communication and architecture changes in project releases.

**Creation time of mailing lists**: From Fig. 5 and Fig. 6, we found that the creation time of the mailing lists from the start of the projects is different between ArgoUML and Hibernate. Developers of ArgoUML created a mailing list to communicate development information before its first version was released (v0.8), which is a common practice in OSS development [22]. Developers of Hibernate created a mailing list after its first version was released (v0.8, which is not shown in Fig. 6). Hibernate initially only had one core developer (GK), and he did not require a mailing list to communicate with other developers before he released the first version to the OSS community. Architectural posts were few when Hibernate v0.9 was released (the time when the mailing list was created). The number of architectural posts increased before Hibernate v1.0 was released because the core developer needed to clarify the implicit architecture information to share with other developers.

**Architecture information communication and architecture changes**: From Fig. 5 and Fig. 6, we found that in ArgoUML, a large proportion (65.2%) of architecture information communication took place before the first stable release (v0.10) because the changes in the architecture of ArgoUML is comparatively stable after the first stable release by following an architecture pattern. On the other hand, Hibernate had more releases than ArgoUML and a large part of architectural discussions in posts (88.4%) was evenly distributed over the time period after the first stable release (Hibernate v1.0.0 final), which is due to the fact that the architecture of Hibernate was refactored over several versions after the first stable release. This refactoring involves identifying new subsystems (e.g., Hibernate Core, Hibernate OGM[18], and Hibernate Search[19]), and the architectural discussions in Hibernate after the first stable version were mainly about these new subsystems.

---

[18] http://in.relation.to/2011/06/17/hibernate-ogm-birth-announcement/
[19] http://in.relation.to/2006/12/20/hibernate-search-hibernate-apache-lucene-integration/

**Architecture information communication and releases of projects**: From Fig. 8 and Fig. 9, we see that the architecture information communication was about system level design (20.5% posts in ArgoUML and 3.8% posts in Hibernate) before the first stable release, and the architecture information communication was mainly about class level design (7.5% posts in ArgoUML and 10.7% posts in Hibernate) after the first stable release. This is reasonable as the focus of architecture design normally starts from the high (system) level, and shifts to a more detailed design level when the architecture is refined, implemented, and refactored. This study shows that there are differences in the timing of architecture design and detail design activities in different OSS projects, probably due to some external contexts related to each project. Developers would be well placed if they *understand the variations between design activities and releases* of each project.

> **What**: Architecture Rationale and Architecture Model are the most frequently communicated architecture information, which is a rich source of architecture documentation.

**Frequently communicated architecture elements**: As reported in Section 4.4, architecture information elements that are most frequently communicated in the mailing lists are Architecture Rationale and Architecture Model. Architecture Rationale element is a natural product of architecture negotiation and interpretation, which are the main purposes of architecture information communication (see Section 4.1). Architecture Rationale and Architecture Model elements are most frequently communicated between developers for constructing design and making decisions [35][38], and developer mailing lists provide a major vehicle for the communication of various knowledge (including architecture knowledge) in OSS development. We did not find any explicit discussions about the architecture elements Architecture View, Architecture Viewpoint, Correspondence Rule, Correspondence, and Model Kind in the developer mailing lists of the two OSS projects. Architecture models were discussed but without explicitly mentioning the architecture views of the models. Certain architecture view was implicitly discussed such as scenarios for describing the quality considerations of stakeholders, and we coded that as Concern element. This result is consistent with the result in our recent survey on architecture documentation in OSS development [16], i.e., there are few architecture views (7.4%) and no architecture viewpoints (0.0%) in the architecture documents of 2,000 OSS projects we investigated.

**Architecture rationale communication without documentation**: As shown in Fig. 7, Architecture Rationale is the most frequently communicated architecture element in the developer mailing lists of both projects, while in our recent survey on OSS SA documentation, architecture rationale is seldom recorded in SA documents of OSS projects [16]. For example, ArgoUML mentions the use of the MVC pattern (i.e., model, view, and control subsystems) in its SA document[20], but the architecture rationale of using MVC is only explained in the mailing list (see the previous paragraph of this section). Similarly, Hibernate provides a high level architecture in objects in its SA document[21], e.g., *Session*, and the architecture rationale of using *Session* is only explained in the mailing list[22]. Developers tend to document the results of architecture design with less focus on the rationale of design due to the cost, effort, and willingness [50]. One suggestion for practitioners on architecture documentation, especially documentation of architecture rationale is to provide lightweight approaches for them, for example, linking the existing discussion of architecture rationale to the architecture documentation [61]. This study has shown that *mailing list is one of the best places to find architecture rationale.*

**Frequently communicated architecture content**: The frequently communicated architecture content is "Adaptive environment" in ArgoUML and "New component" in Hibernate. The reasons are the following: ArgoUML is a GUI-based tool for UML modeling, and most architecture information communication concerns the GUI requirements of the system, e.g., discussions about how to decouple the GUI from the logic layer when introducing the feature of multiple users[23], and Hibernate introduces many new components when extending from a core system to several subsystems. "Modularity" is the most communicated internal quality requirement and "Extensibility" is the most communicated external quality requirement in ArgoUML, because ArgoUML follows "*a modular, extendable structure*"[24] based on the MVC pattern in its architecture design. This explanation is supported by Harrison *et al.* that "*extensibility and maintainability were cited as motivations for employing the MVC pattern*" in OSS design [24]. "Modularity" is also the most communicated internal quality requirement in Hibernate, which is not a surprising result, because Hibernate evolves from a core system to several subsystems as mentioned before, and "Modularity" is a critical concern in this decomposition. "Efficiency" is the most communicated external quality requirement in Hibernate, because Hibernate provides a framework for mapping an object-oriented data model (e.g., Java classes) to a relational database (e.g., database tables), and "Efficiency" is an important internal quality requirement for using database systems.

---

[20] http://argouml-stats.tigris.org/documentation/defaulthtml/cookbook/ch04.html
[21] http://docs.jboss.org/hibernate/core/4.3/manual/en-US/html/ch02.html
[22] http://lists.jboss.org/pipermail/hibernate-dev/2014-March/011110.html
[23] http://argouml.tigris.org/ds/viewMessage.do?dsForumId=450&dsMessageId=210031
[24] http://argouml.tigris.org/ds/viewMessage.do?dsForumId=450&viewType=browseAll&dsMessageId=116987

## 5.2 Implications for OSS Architecting

**More than one architect in OSS projects**: This finding confirms the claim by Bass *et al.* in [8] that "*most OSS projects do not have a single identifiable architect, and the architecture is typically the shared responsibility of the group of committers*". Using social network analysis, we found in this study that a few developers intensively communicate architectural information, and they play the role of architect in OSS development. Research shows that there is a correlation between the number of developers who work on an OSS project and the age of the project [37]. With the increasing number of developers in OSS projects, multiple developers with architectural inputs in OSS projects may be the norm.

We observe that architecture information is negotiated and decided between a few core developers. The natural design organization tendency seems to be having fewer developers, probably who are more knowledgeable and involved in the project, to make architectural decisions. Despite the possibility that all developers in OSS development can participate as they please and their roles are not assigned, we can see a centrality tendency for a few developers to take on the role of architects. This group of people fulfil the duties of architecting and communicating amongst developers as described in [52].

**Architecture information communication and architecture evolution**: Architecture evolution has a significant impact on the whole system [53]. Our results indicate that there is a relationship between the timing of architecture information communication and how the architecture stabilizes. For example, the amount of architectural threads communicated before the first stable release in ArgoUML is 65.2% and after the first stable release in Hibernate is 88.4%. In our previous work [15], we found that 89.1% architecture changes happened before the first stable release in ArgoUML and 54.3% architecture changes happened after the first stable release in Hibernate, which is positively related to the results of RQ3. Architecture information communication precedes architecture changes, and consequently the amount of architecture information communication indicates the frequency of architecture changes. Collecting architecture information communication could help to identify and understand the changes of architectures.

**Extracting architecture rationale information**: We have found that architecture information recorded in OSS developer mailing lists is rich with design rationale information (the most frequently communicated architecture element as shown in Fig. 7). This type of information is beneficial for enriching architectural decisions [38] in OSS architecture documentation [16]. This information is an important means to enable developers to understand design decisions and communicate design intents. However, there is limited research and research methodologies on studying the behavior and the process of how architects make decisions [54]. Additionally, with the current lack of tools and capabilities, researchers might consider natural language processing techniques for exacting and classifying architecture information (e.g., design rationale information) from OSS developer mailing lists (semi-)automatically [51]. It is also practically meaningful to link posts that contain architecture rationale information to architecture documentation or use extracted architecture rationale information to enrich SA documentation in OSS development.

**Modularity and conciseness requirements**: Modularity and conciseness requirements are the top two most frequently communicated internal quality requirements in both ArgoUML and Hibernate as shown in Fig. 8 and Fig. 9. This finding shows that software modularity and conciseness are critical concerns when designing the architecture of an OSS. In OSS development, generally more than one developer work on the system (e.g., subsystems or components). Such distributed development naturally dictates a system structure such that developers try to avoid frequently modifying or unconsciously violating others' parts of the system. As such, modularity and comprehensible architecture of an OSS system become essential quality requirements.

**Supplementing architecture communication in OSS development**: We analyzed architecture elements communicated in OSS developer mailing lists in order to understand what techniques are employed when OSS developers make architectural decisions. According to the comparative survey of decision making techniques in architecture design [17], the characteristics of decision making technique employed in these two OSS projects can be classified as: *Just a term* (e.g., using just a term for describing a quality attribute), *No articulation* (e.g., the level of desire for a particular quality attribute is not articulated), *On/Off* (e.g., accept or reject whether an alternative fulfills a quality attribute), and *Uncertainty is inferred from disagreement among decision makers* (e.g., decision makers expressing and communicating their opinions about the fulfillment of a decision). This result implies that architecture information communication through mailing lists for making architectural decisions is vague, which may negatively impact the quality of the decisions, and verbal communication or other means should also be employed to facilitate architectural decision making.

**Understanding architectural decisions**: Architecture Rationale is most frequently communicated in OSS developer mailing lists (i.e., over 70% in all architectural posts as shown in Fig. 7) compared with the architecture rationale documented in OSS SA documentation (only 18% in all OSS projects have SA documentation) [16]. This result implies that OSS developers are more likely to discuss architecture rationale in developer mailing lists, but they seldom document rationale in SA documentation. This result supports Kazman *et al.*'s findings that architectural discussion can facilitate architectural thinking and reasoning in OSS development through understanding architectural decisions [29]. However, design documentation and design rationale contained within developer mailing lists are disjointed and finding a way to connect them could enrich developers' understanding of a design.

**Relating requirements and architecture through architecture negotiation**: From Fig. 10 and Fig. 11 with the related architecture threads, we found that architecture negotiation and interpretation are the main purposes of the communication when architecture topics and quality requirements were discussed. It implies that architects use architecture negotiation and interpretation to try to understand the design and the requirements before making decisions. These two communication techniques pave the way for relating requirements and architecture in architecture design in the early stages of a project. Whilst we know that OSS developers carry out such communication, there have been little real-world studies on how best to achieve communication in the OSS community [54]. This is an area for future study.

To conclude, we provide a summary of the results and the implications of this multiple case study from ArgoUML and Hibernate in Table 6.

Table 6. Summary of the results and the implications of this study

| Research Question | Results and Implications for OSS Architecting |
|---|---|
| RQ1: Purposes of communicating architecture information (Why) | • Negotiation and interpretation of architecture are the main purposes of architecture information communication (see Fig. 1 and Fig. 2), and these two aspects of architecture communication help architects make decisions in OSS development.<br>• Architecture negotiation and interpretation pave the way for relating requirements and architecture in architecture design (see Fig. 10 and Fig. 11). |
| RQ2: Social network of architecture information communication (Who) | • Architecture information is mainly communicated between few core developers who make architectural decisions (see Fig. 3 and Fig. 4).<br>• Being aware of the core developers would help to identify and collect more architectural decisions and implement an effective personalization strategy in architectural knowledge management.<br>• The core developers of architecture information communication can change between various release versions. |
| RQ3: Architectural posts and threads publishing timing (When) | • There is a relationship between the timing of architecture information communication (see Fig. 5, Fig. 6 and Fig. 8, Fig. 9) and architecture changes in project releases.<br>• Collecting architecture information communication could help to identify and understand the changes of architectures. |
| RQ4: Architecture information communicated during development (What) | • Architecture Rationale and Architecture Model are most frequently communicated in the developer mailing lists (see Fig. 7).<br>• The discussion on rationale of architecture design can be used to enrich architecture documentation in OSS development.<br>• Architecture topics and quality requirements are extensively discussed together in both projects (see Fig. 10 and Fig. 11).<br>• The characteristics of decision making technique employed in mailing lists (e.g., *Just a term*) imply that architecture information communication for making architectural decisions is vague, and other complementary means should be employed to facilititate architecture communication. |

# 6. Threats to validity

We discuss the threats to the validity of this study by following the guidelines in [23]. Internal validity is not considered because we did not study causality and used descriptive statistics only to describe the results.

**Construct validity** denotes whether the measures used in a study can represent the constructs in the real setting. The threat to this validity in our study is whether architectural information was extracted correctly. To mitigate this threat, the commonly accepted conceptual model for architectural description in the ISO 42010:2011 standard [25] was used to help us judge whether a post in mailing lists contains architecture information. But using a standard cannot guarantee the correct extraction of architectural information because different researchers may understand the architecture elements in the conceptual model differently. To mitigate this issue, three researchers conducted the pilot and formal data extraction and identification of the architectural posts and elements, and any posts which the first two authors could not decide their types, were further discussed and confirmed with the third author. Any disagreements on the architecture elements in the architectural posts and the coding results of the content of architecture information communication were discussed and resolved by all the authors. These measures were used to reduce personal bias and increase the correctness of extracting architectural posts, elements, and codes in mailing lists.

**External validity** concerns the generality of the study. We have selected two sizeable OSS projects that have mailing lists and with the selection criteria described in Section 3.3. The two selected projects are from different domains. Both projects are of decent size and duration (over 18 years). However, the results may not be applicable to other OSS projects that have shorter development periods or that were started in the last five years. Our current data appears to support certain observations made in other studies (e.g., in [29][30]). We acknowledge that whilst our evidence appears to be consistent with these observations, our claims are limited to our dataset. The use of this systematic approach based on encoding within Grounded Theory to analyze architecture communication can be employed to analyze communication data from other sources (e.g., issue tracking systems). Another threat is whether the data source (i.e., developer mailing lists) we chose is suitable for architectural information collection and analysis. To reduce this threat, we manually checked other channels for knowledge sharing and communication in these two OSS projects, and we found that in these two OSS projects, the developers mainly used mailing lists for communication. A recent study on 37 Apache projects also confirms that most of the design discussions (89.51%) occur in project mailing lists [62]. However, it is still possible that some architecture information was communicated through other channels or recorded in documentation. More data sources from various communication channels (e.g., issue trackers and Wikis) can be included for a better and more comprehensive analysis, which will be considered in our future work.

**Reliability** refers to whether the study gets the same results when other researchers repeat it. The threats to the reliability of this study concern the collection and analysis processes of architectural information communicated in OSS developer mailing lists. We made explicit the processes of how to collect and analyze data in this study (as detailed in Section 3.4 and Section 3.5, respectively), and conducted the data collection and analysis in several iterations. We also employed a systematic encoding approach to manually analyze the qualitative data in this work, to partially improve the reproducibility of the analysis results. These measures partially mitigated the threats.

## 7. Conclusions and future work

We report a multiple case study that characterizes why, who, when, and what OSS developers use mailing lists to communicate architecture information in OSS development. We conducted this investigation by extracting architecture information from the developer mailing lists of two popular OSS projects: ArgoUML and Hibernate. We found a few similarities of architecture communication between the two OSS projects: the purposes of architecture communication in the two projects are similar; the timing and amount of architecture information communicated in developer mailing lists of the two projects are similar, they decreased significantly after the first stable releases; architecture information communication mostly took place amongst few core developers; and the communicated architectural elements are also similar. These similarities reveal how OSS developers naturally gravitate towards the four architecture communication aspects in OSS development (i.e., the Why, Who, When, and What).

This study finds that the amount of architecture information communication can indicate the frequency of architecture changes. Architectural information communicated in developer mailing lists (e.g., rationale, model, and concern) can be extracted and used to enrich SA documentation, especially the documentation of architectural design decisions, in OSS development.

This study has initiated a wide range of opportunities to study OSS development communications. We see several promising research directions: (1) Other sources of architecture information such as, issue tracking systems, forums, commit data [10][26], twitter, and blogs [34], may also be analyzed in OSS development to study architecture communication. We plan to research how architecture information is communicated and shared in those sources, and identify the differences (as well as their benefits and weakness) compared to architecture information communication using mailing lists. It is worthwhile to study OSS communication characteristics by including more OSS projects in the investigation. We want to study architecture communication in different project domains, and different project sizes etc. in order to generalize the results. (2) Study the impact of architecture information communication on architecture changes, e.g., relationships or patterns between the architecture information communicated and architecture changes recorded in patches or commit data. (3) Study the causes that lead to architecture changes according to the communication topics submitted to issue tracking systems. (4) Develop (semi)-automatic tools to extract architecture information communicated in developer mailing lists to enhance project documentation. (5) Conduct surveys and interviews on how developers perceive architecture information communication in practice, and then combine developers' opinions with the results of this study to develop guidelines for architecture communication and documentation in the OSS community.

## Acknowledgments

This work is partially sponsored by the National Key R&D Program of China with Grant No. 2018YFB1402800. We would also like to thank Hans van Vliet for his valuable suggestions on an early draft of this work.